\documentclass[prd,onecolumn,footinbib,showpacs]{revtex4}
\usepackage{amsfonts}
\newcommand\gothfamily{\usefont{U}{ygoth}{m}{n}}
\DeclareTextFontCommand{\textgoth}{\gothfamily}

\begin{document}

\title{A unified, purely affine theory of gravitation and electromagnetism} 
\author{Nikodem J. Pop\l awski}
\affiliation{Department of Physics, Indiana University,
727 East Third Street, Bloomington, Indiana 47405, USA}
\date{\today}

\begin{abstract}
In the purely affine formulation of gravity, the gravitational field is represented by the symmetric part of the Ricci tensor of the affine connection.
The classical electromagnetic field can be represented in this formulation by the second Ricci tensor of the connection.
Such a construction is dynamically equivalent to the sourceless Einstein--Maxwell equations.
We generalize this construction to the case with sources, represented by the derivative of the affine Lagrangian density with respect to the connection.
We show that the Maxwell equations with sources emerge for the simplest affine Lagrangian for matter, while the Einstein and Lorentz equations arise if mass has electromagnetic origin.
We also show that the Maxwell equations replace the unphysical constraint imposed by the projective invariance of purely affine Lagrangians that depend explicitly on the connection.
\end{abstract}
\pacs{04.50.+h, 04.20.Fy, 03.50.-z}

\maketitle
\setcounter{footnote}{0}

In the {\em purely affine} (Einstein--Eddington) formulation of general relativity~\cite{Ein,Edd,Schr1,Schr2,Kij,CFK}, a Lagrangian density depends on a torsionless affine connection and the symmetric part of the Ricci tensor of the connection.
This formulation defines the metric tensor as the derivative of the Lagrangian density with respect to the Ricci tensor, obtaining an algebraic relation between these two tensors.
It derives the field equations by varying the total action with respect to the connection, which gives a differential relation between the connection and the metric tensor.
This relation yields a differential equation for the metric.
In the {\em metric--affine} (Einstein--Palatini) formulation~\cite{Ein,Pal}, both the metric tensor and the torsionless connection are independent variables, and the field equations are derived by varying the action with respect to these quantities.
The corresponding Lagrangian density is linear in the symmetric part of the Ricci tensor of the connection.
In the {\em purely metric} (Einstein--Hilbert) formulation~\cite{Hilb,LL2,MTW}, the metric tensor is a variable, the affine connection is the Levi-Civita connection of the metric and the field equations are derived by varying the action with respect to the metric tensor.
The corresponding Lagrangian density is linear in the symmetric part of the Ricci tensor of the metric.
All three formulations of general relativity are dynamically equivalent~\cite{FK3}.
This statement can be generalized to theories of gravitation with Lagrangians that depend on the full Ricci tensor and the second Ricci tensor~\cite{univ,nonsym}, and to a general connection with torsion~\cite{nonsym}.

The purely affine formulation of gravity allows an elegant unification of the classical free electromagnetic and gravitational fields.
Ferraris and Kijowski showed that the gravitational field is represented by the symmetric part of the Ricci tensor of the connection (not restricted to be symmetric), while the electromagnetic field can be represented by the second Ricci tensor of the connection~\cite{FK2}.
Such a construction is dynamically equivalent to the sourceless Einstein--Maxwell equations~\cite{FK2}.
In this paper, we generalize this construction to the case with sources, represented by the derivative of the affine Lagrangian density with respect to the connection.
We show that the Maxwell equations with sources emerge in this unified framework for the simplest affine Lagrangian for matter, while the Einstein and Lorentz equations arise if mass has electromagnetic origin.
We also show that the Maxwell equations replace the unphysical constraint imposed by the projective invariance of purely affine Lagrangians.

A general purely affine Lagrangian density $\textgoth{L}$ depends on the affine connection $\Gamma^{\,\,\rho}_{\mu\,\nu}$ and the curvature tensor, $R^\rho_{\phantom{\rho}\mu\sigma\nu}=\Gamma^{\,\,\rho}_{\mu\,\nu,\sigma}-\Gamma^{\,\,\rho}_{\mu\,\sigma,\nu}+\Gamma^{\,\,\kappa}_{\mu\,\nu}\Gamma^{\,\,\rho}_{\kappa\,\sigma}-\Gamma^{\,\,\kappa}_{\mu\,\sigma}\Gamma^{\,\,\rho}_{\kappa\,\nu}$.
Let us assume that the dependence of the Lagrangian on the curvature is restricted to the contracted curvature tensors~\cite{nonsym}, of which there exist three: the symmetric $P_{\mu\nu}=R_{(\mu\nu)}$ and antisymmetric $R_{[\mu\nu]}$ part of the Ricci tensor, $R_{\mu\nu}=R^\rho_{\phantom{\rho}\mu\rho\nu}$, and the antisymmetric second Ricci tensor, $Q_{\mu\nu}=R^\rho_{\phantom{\rho}\rho\mu\nu}=\Gamma^{\,\,\rho}_{\rho\,\nu,\mu}-\Gamma^{\,\,\rho}_{\rho\,\mu,\nu}$, which has the form of a curl~\cite{Schr2,Scho}.
Accordingly, the variation of the corresponding action is given by $\delta S=\frac{1}{c}\delta\int d^4x\textgoth{L}(\Gamma,R,Q)=\frac{1}{c}\int d^4x\Bigl(\frac{\partial\textgoth{L}}{\delta\Gamma^{\,\,\rho}_{\mu\,\nu}}\delta\Gamma^{\,\,\rho}_{\mu\,\nu}+\frac{\partial\textgoth{L}}{\delta R_{\mu\nu}}\delta R_{\mu\nu}+\frac{\partial\textgoth{L}}{\delta Q_{\mu\nu}}\delta Q_{\mu\nu}\Bigr)$.

The metric structure associated with a purely affine Lagrangian is obtained using~\cite{Edd,Schr2,Niko1}:
\begin{equation}
{\sf g}^{\mu\nu}\equiv-2\kappa\frac{\partial\textgoth{L}}{\partial R_{\mu\nu}},
\label{met1}
\end{equation}
where ${\sf g}^{\mu\nu}$ is the fundamental tensor density and $\kappa=\frac{8\pi G}{c^4}$ (for purely affine Lagrangians that do not depend on $R_{[\mu\nu]}$ this definition is equivalent to that in~\cite{Kij,FK3,FK2,FK1}: ${\sf g}^{\mu\nu}\equiv-2\kappa\frac{\partial\textgoth{L}}{\partial P_{\mu\nu}}$).
The contravariant metric tensor is defined by
\begin{equation}
g^{\mu\nu}\equiv\frac{{\sf g}^{(\mu\nu)}}{\sqrt{-\mbox{det}{\sf g}^{(\rho\sigma)}}}.
\label{met2}
\end{equation}
To make this definition meaningful, we have to assume $\mbox{det}({\sf g}^{(\mu\nu)})\neq0$, which also guarantees that the tensor $g^{\mu\nu}$ has the Lorentzian signature $(+,-,-,-)$~\cite{FK3}.
The covariant metric tensor $g_{\mu\nu}$ is related to the contravariant metric tensor by $g^{\mu\nu}g_{\rho\nu}=\delta^\mu_\rho$.
The tensors $g^{\mu\nu}$ and $g_{\mu\nu}$ are used for raising and lowering indices.
We also define an antisymmetric tensor density:
\begin{equation}
{\sf h}^{\mu\nu}\equiv-2\kappa\frac{\partial\textgoth{L}}{\partial Q_{\mu\nu}},
\label{amet1}
\end{equation}
and the density conjugate to the connection:
\begin{equation}
\Pi_{\phantom{\mu}\rho\phantom{\nu}}^{\mu\phantom{\rho}\nu}\equiv-2\kappa\frac{\partial\textgoth{L}}{\partial \Gamma^{\,\,\rho}_{\mu\,\nu}},
\label{con1}
\end{equation}
which has the same dimension as the connection.
Consequently, the variation of the action can be written as $\delta S=-\frac{1}{2\kappa c}\int d^4x(\Pi_{\phantom{\mu}\rho\phantom{\nu}}^{\mu\phantom{\rho}\nu}\delta\Gamma^{\,\,\rho}_{\mu\,\nu}+{\sf g}^{\mu\nu}\delta P_{\mu\nu}+{\sf h}^{\mu\nu}\delta Q_{\mu\nu})$.

If we do not restrict the connection $\Gamma^{\,\,\rho}_{\mu\,\nu}$ to be symmetric, the variation of the Ricci tensor is given by the Palatini formula~\cite{Schr2}: $\delta R_{\mu\nu}=\delta\Gamma^{\,\,\rho}_{\mu\,\nu;\rho}-\delta\Gamma^{\,\,\rho}_{\mu\,\rho;\nu}-2S^\sigma_{\phantom{\sigma}\rho\nu}\delta\Gamma^{\,\,\rho}_{\mu\,\sigma}$, where $S^\rho_{\phantom{\rho}\mu\nu}=\Gamma^{\,\,\,\,\rho}_{[\mu\,\nu]}$ is the torsion tensor and the semicolon denotes the covariant differentiation with respect to $\Gamma^{\,\,\rho}_{\mu\,\nu}$.
Using the identity $\int d^4x({\sf V}^\mu)_{;\mu}=2\int d^4x S_\mu{\sf V}^\mu$, where ${\sf V}^\mu$ is an arbitrary vector density and $S_\mu=S^\nu_{\phantom{\nu}\mu\nu}$ is the torsion vector~\cite{Schr2}, and applying the principle of least action $\delta S=0$, we obtain
\begin{equation}
{\sf g}^{\mu\nu}_{\phantom{\mu\nu};\rho}-{\sf g}^{\mu\sigma}_{\phantom{\mu\sigma};\sigma}\delta^\nu_\rho-2{\sf g}^{\mu\nu}S_\rho+2{\sf g}^{\mu\sigma}S_\sigma\delta^\nu_\rho+2{\sf g}^{\mu\sigma}S^\nu_{\phantom{\nu}\rho\sigma}=\Pi_{\phantom{\mu}\rho\phantom{\nu}}^{\mu\phantom{\rho}\nu}+2{\sf h}^{\nu\sigma}_{\phantom{\nu\sigma},\sigma}\delta^\mu_\rho.
\label{field1}
\end{equation}
This equation is equivalent to
\begin{equation}
{\sf g}^{\mu\nu}_{\phantom{\mu\nu},\rho}+\,^\ast\Gamma^{\,\,\mu}_{\sigma\,\rho}{\sf g}^{\sigma\nu}+\,^\ast\Gamma^{\,\,\nu}_{\rho\,\sigma}{\sf g}^{\mu\sigma}-\,^\ast\Gamma^{\,\,\sigma}_{\sigma\,\rho}{\sf g}^{\mu\nu}=\Pi_{\phantom{\mu}\rho\phantom{\nu}}^{\mu\phantom{\rho}\nu}-\frac{1}{3}\Pi_{\phantom{\mu}\sigma\phantom{\sigma}}^{\mu\phantom{\sigma}\sigma}\delta^\nu_\rho+2{\sf h}^{\nu\sigma}_{\phantom{\nu\sigma},\sigma}\delta^\mu_\rho-\frac{2}{3}{\sf h}^{\mu\sigma}_{\phantom{\mu\sigma},\sigma}\delta^\nu_\rho,
\label{field2}
\end{equation}
where $^\ast\Gamma^{\,\,\rho}_{\mu\,\nu}=\Gamma^{\,\,\rho}_{\mu\,\nu}+\frac{2}{3}\delta^\rho_\mu S_\nu$~\cite{Schr1,Schr2}.

Contracting the indices $\mu$ and $\rho$ in Eq.~(\ref{field1}) yields
\begin{equation}
{\sf g}^{[\nu\sigma]}_{\phantom{[\nu\sigma]},\sigma}+\frac{1}{2}\Pi_{\phantom{\sigma}\sigma\phantom{\nu}}^{\sigma\phantom{\sigma}\nu}+4{\sf h}^{\nu\sigma}_{\phantom{\nu\sigma},\sigma}=0,
\label{Max1}
\end{equation}
which generalizes one of the field equations of Schr\"{o}dinger's purely affine gravity with the nonsymmetric metric tensor~\cite{Schr1,Schr2}.
Let us assume that the Lagrangian density $\textgoth{L}$ does not depend on $R_{[\mu\nu]}$.
As a result, we have $\frac{\partial\textgoth{L}}{\partial R_{\mu\nu}}=\frac{\partial\textgoth{L}}{\partial P_{\mu\nu}}$.
Consequently, the tensor density ${\sf g}^{\mu\nu}$ is symmetric and Eq.~(\ref{Max1}) reduces to
\begin{equation}
{\sf h}^{\sigma\nu}_{\phantom{\sigma\nu},\sigma}={\sf j}^\nu\equiv\frac{1}{8}\Pi_{\phantom{\sigma}\sigma\phantom{\nu}}^{\sigma\phantom{\sigma}\nu},
\label{Max2}
\end{equation}
which has the form of the Maxwell equations for the electromagnetic field.
The density $\Pi_{\phantom{\mu}\rho\phantom{\nu}}^{\mu\phantom{\rho}\nu}$ represents the {\em source} for the purely affine field equations.
Since the tensor density ${\sf h}^{\mu\nu}$ is antisymmetric, the current vector density ${\sf j}^\mu$ must be conserved: ${\sf j}^\mu_{\phantom{\mu},\mu}=0$, which constrains how the connection $\Gamma^{\,\,\rho}_{\mu\,\nu}$ can enter a purely affine Lagrangian density $\textgoth{L}$: $\Pi_{\phantom{\sigma}\sigma\phantom{\nu},\nu}^{\sigma\phantom{\sigma}\nu}=0$.
We note that this conservation is valid even if $\textgoth{L}$ depends on $R_{[\mu\nu]}$.
We also note that if $\textgoth{L}$ does not depend on $Q_{\mu\nu}$, the field equation~(\ref{Max2}) becomes a stronger, algebraic constraint on how the Lagrangian depends on the connection: $\Pi_{\phantom{\sigma}\sigma\phantom{\nu}}^{\sigma\phantom{\sigma}\nu}=0$.
The dependence of a purely affine Lagrangian on the second Ricci tensor $Q_{\mu\nu}$, which will be associated with the electromagnetic field, {\em replaces} this unphysical constraint with a field equation for the tensor density ${\sf h}^{\mu\nu}$ (the Maxwell equations).

Another way to overcome the constraint $\Pi_{\phantom{\sigma}\sigma\phantom{\nu}}^{\sigma\phantom{\sigma}\nu}=0$ is to restrict the torsion tensor to be traceless: $S_\mu=0$~\cite{San}.
This condition enters the Lagrangian density as a Lagrange multiplier term $-\frac{1}{2\kappa}{\sf B}^\mu S_\mu$, where the Lagrange multiplier ${\sf B}^\mu$ is a vector density.
Consequently, there is an extra term ${\sf B}^{[\mu}\delta^{\nu]}_\rho$ on the right-hand side of Eq.~(\ref{field1}) and Eq.~(\ref{Max2}) becomes ${\sf h}^{\sigma\nu}_{\phantom{\sigma\nu},\sigma}={\sf j}^\nu-\frac{3}{16}{\sf B}^\nu$.
Setting ${\sf B}^\nu=\frac{16}{3}{\sf j}^\nu$ removes this constraint if $\textgoth{L}$ does not depend on $Q_{\mu\nu}$, or yields the wave equation ${\sf h}^{\sigma\nu}_{\phantom{\sigma\nu},\sigma}=0$ if $\textgoth{L}$ depends on $Q_{\mu\nu}$.
In both cases, the vector density ${\sf j}^\mu$ does not need to be conserved.
Therefore introducing the dependence of $\textgoth{L}$ on $Q_{\mu\nu}$ is more suitable for unifying gravitation and electromagnetism in a purely affine formalism than imposing $S_\mu=0$.

The tensor $P_{\mu\nu}$ is invariant under a projective transformation $\Gamma^{\,\,\rho}_{\mu\,\nu}\rightarrow\Gamma^{\,\,\rho}_{\mu\,\nu}+\delta^\rho_\mu V_\nu$ (or $\delta\Gamma^{\,\,\rho}_{\mu\,\nu}=\delta^\rho_\mu\delta V_\nu$), and so is ${\sf g}^{\mu\nu}$.
Under the same transformation, the tensor $Q_{\mu\nu}$ changes according to $Q_{\mu\nu}\rightarrow Q_{\mu\nu}+4(V_{\nu,\mu}-V_{\mu,\nu})$.
Consequently, the action changes according to $\delta S=-\frac{1}{2\kappa c}\int d^4x(\Pi_{\phantom{\mu}\rho\phantom{\nu}}^{\mu\phantom{\rho}\nu}\delta\Gamma^{\,\,\rho}_{\mu\,\nu}+{\sf h}^{\mu\nu}\delta Q_{\mu\nu})=-\frac{1}{2\kappa c}\int d^4x(\Pi_{\phantom{\sigma}\sigma\phantom{\mu}}^{\sigma\phantom{\sigma}\mu}+8{\sf h}^{\mu\nu}_{\phantom{\mu\nu},\nu})\delta V_\mu$.
This expression is identically zero due to the field equation~(\ref{Max2}) so the action is projective invariant.
We can interpret the electromagnetic field in a purely affine gravity as a field whose role is to {\em preserve} the projective invariance of purely affine Lagrangians that depend explicitly on the affine connection without constraining the connection.

If we apply to $\textgoth{L}(\Gamma,P,Q)$ the Legendre transformation with respect to $P_{\mu\nu}$~\cite{Kij,FK3}, defining the Hamiltonian density $\textgoth{H}$:
\begin{equation}
\textgoth{H}=\textgoth{L}-\frac{\partial\textgoth{L}}{\partial P_{\mu\nu}}P_{\mu\nu}=\textgoth{L}+\frac{1}{2\kappa}{\sf g}^{\mu\nu}P_{\mu\nu},
\label{Leg1}
\end{equation}
we find: $d\textgoth{H}=\frac{\partial\textgoth{L}}{\partial \Gamma^{\,\,\rho}_{\mu\,\nu}}d\Gamma^{\,\,\rho}_{\mu\,\nu}+\frac{1}{2\kappa}P_{\mu\nu}d{\sf g}^{\mu\nu}+\frac{\partial\textgoth{L}}{\partial Q_{\mu\nu}}dQ_{\mu\nu}$.
Accordingly, the Hamiltonian density $\textgoth{H}$ is a function of $\Gamma^{\,\,\rho}_{\mu\,\nu}$, ${\sf g}^{\mu\nu}$ and $Q_{\mu\nu}$, and the action variation is: $\delta S=\frac{1}{c}\delta\int\Bigl(\textgoth{H}(\Gamma,{\sf g},Q)-\frac{1}{2\kappa}{\sf g}^{\mu\nu}P_{\mu\nu}(\Gamma)\Bigr)d^4x=\frac{1}{c}\int\Bigl(\frac{\partial\textgoth{H}}{\partial\Gamma^{\,\,\rho}_{\mu\,\nu}}\delta\Gamma^{\,\,\rho}_{\mu\,\nu}+\frac{\partial\textgoth{H}}{\partial{\sf g}^{\mu\nu}}\delta{\sf g}^{\mu\nu}-\frac{1}{2\kappa}{\sf h}^{\mu\nu}\delta Q_{\mu\nu}-\frac{1}{2\kappa}{\sf g}^{\mu\nu}\delta P_{\mu\nu}-\frac{1}{2\kappa}P_{\mu\nu}\delta{\sf g}^{\mu\nu}\Bigr)d^4x$.
The variation with respect to ${\sf g}^{\mu\nu}$ yields the first Hamilton equation~\cite{Kij,FK3}:
\begin{equation}
P_{\mu\nu}=2\kappa\frac{\partial\textgoth{H}}{\partial {\sf g}^{\mu\nu}}.
\label{Ham1}
\end{equation}
The variations with respect to $P_{\mu\nu}$ and $Q_{\mu\nu}$ can be transformed to the variation with respect to $\Gamma^{\,\,\rho}_{\mu\,\nu}$ by means of the Palatini formula and the variation of a curl, respectively, giving the second Hamilton equation equivalent to the field equations~(\ref{field1}).

The analogous transformation in classical mechanics goes from a Lagrangian $L(q^i,\dot{q}^i)$ to a Hamiltonian $H(q^i,p^i)=p^j\dot{q}^j-L(q^i,\dot{q}^i)$ (or, more precisely, a Routhian since not all the variables are subject to a Legendre transformation~\cite{LL1}) with $p^i=\frac{\partial{L}}{\partial\dot{q}^i}$, where the tensor $P_{\mu\nu}$ corresponds to {\em generalized velocities} $\dot{q}^i$ and the density ${\sf g}^{\mu\nu}$ to {\em canonical momenta} $p^i$~\cite{Kij,FK3}.
Accordingly, the affine connection plays the role of the {\em configuration} $q^i$ and the source density $\Pi_{\phantom{\mu}\rho\phantom{\nu}}^{\mu\phantom{\rho}\nu}$ corresponds to {\em generalized forces} $f^i=\frac{\partial{L}}{\partial q^i}$~\cite{Kij}.
The field equations~(\ref{field2}) correspond to the Lagrange equations $\frac{\partial L}{\partial q^i}=\frac{d}{dt}\frac{\partial L}{\partial\dot{q}^i}$ which result from Hamilton's principle $\delta\int L(q^i,\dot{q}^i)dt=0$ for arbitrary variations $\delta q^i$ vanishing at the boundaries of the configuration.
The Hamilton equations result from the same principle written as $\delta\int(p^j\dot{q}^j-H(q^i,p^i))dt=0$ for arbitrary variations $\delta q^i$ and $\delta p^i$~\cite{LL1}.
The field equations~(\ref{field1}) correspond to the second Hamilton equation, $\dot{p}^i=-\frac{\partial H}{\partial q^i}$, and Eq.~(\ref{Ham1}) to the first Hamilton equation, $\dot{q}^i=\frac{\partial H}{\partial p^i}$.

From Eq.~(\ref{Ham1}) it follows that $2\kappa\delta\textgoth{H}=P_{\mu\nu}\delta{\sf g}^{\mu\nu}=(P_{\mu\nu}-\frac{1}{2}Pg_{\mu\nu})\sqrt{-g}\delta g^{\mu\nu}$, where $P=P_{\mu\nu}g^{\mu\nu}$ and $g=\mbox{det}g_{\mu\nu}$.
This expression has the form of the Einstein equations of general relativity:
\begin{equation}
P_{\mu\nu}(\Gamma)-\frac{1}{2}P(\Gamma)g_{\mu\nu}=\kappa\Theta_{\mu\nu},
\label{Ein}
\end{equation}
if we identify $\textgoth{H}$ with the Lagrangian density for matter $\mathcal{L}_{ma}$ in the metric--affine formulation of gravitation, since the symmetric energy--momentum tensor $\Theta_{\mu\nu}$ is defined by the variational relation: $2\kappa\delta\mathcal{L}_{ma}=\Theta_{\mu\nu}\delta{\sf g}^{\mu\nu}$.
From Eq.~(\ref{Leg1}) it follows that $-\frac{1}{2\kappa}P(\Gamma)\sqrt{-g}$ is the metric--affine Lagrangian density for the gravitational field $\mathcal{L}_g$, in agreement with the general-relativistic form.
The transition from the affine to the metric--affine formalism shows that the gravitational Lagrangian density $\mathcal{L}_g$ is a {\em Legendre term} corresponding to $p^j\dot{q}^j$ in classical mechanics~\cite{Kij}.
Therefore the purely affine and metric--affine formulation of gravitation are dynamically equivalent if $\textgoth{L}$ depends on the affine connection, the symmetric part of the Ricci tensor~\cite{FK3} and the second Ricci tensor~\cite{FK2}.
We also note that the metric--affine Lagrangian density for the gravitational field $\mathcal{L}_g$ automatically turns out to be linear in the curvature tensor.
Thus metric--affine Lagrangians for the gravitational field that are nonlinear with respect to curvature {\em cannot} be derived from a purely affine Lagrangian that depends on the connection and the contracted curvature tensors.

Substituting Eq.~(\ref{Max2}) to Eq.~(\ref{field2}) and symmetrizing the indices $\mu$ and $\nu$ yield
\begin{equation}
{\sf g}^{\mu\nu}_{\phantom{\mu\nu},\rho}+\,^\ast\Gamma^{\,\,\,\,\mu}_{(\sigma\,\rho)}{\sf g}^{\sigma\nu}+\,^\ast\Gamma^{\,\,\,\,\nu}_{(\rho\,\sigma)}{\sf g}^{\mu\sigma}-\,^\ast\Gamma^{\,\,\,\,\sigma}_{(\sigma\,\rho)}{\sf g}^{\mu\nu}=\Sigma_{\phantom{\mu}\rho\phantom{\nu}}^{\mu\phantom{\rho}\nu},
\label{field3}
\end{equation}
where
\begin{equation}
\Sigma_{\phantom{\mu}\rho\phantom{\nu}}^{\mu\phantom{\rho}\nu}=\Pi_{\phantom{(\mu}\rho\phantom{\nu)}}^{(\mu\phantom{\rho}\nu)}-\frac{1}{3}\delta^{(\mu}_\rho\Pi_{\phantom{\nu)}\sigma\phantom{\sigma}}^{\nu)\phantom{\sigma}\sigma}-\frac{1}{6}\Pi_{\phantom{\sigma}\sigma\phantom{(\mu}}^{\sigma\phantom{\sigma}(\mu}\delta^{\nu)}_\rho.
\end{equation}
Eq.~(\ref{field3}) is a linear algebraic equation for $^\ast\Gamma^{\,\,\,\,\rho}_{(\mu\,\nu)}$ as a function of the metric tensor, its first derivatives and the density $\Pi_{\phantom{\mu}\rho\phantom{\nu}}^{\mu\phantom{\rho}\nu}$.
We decompose the connection $^\ast\Gamma^{\,\,\rho}_{\mu\,\nu}$ as
\begin{equation}
^\ast\Gamma^{\,\,\rho}_{\mu\,\nu}=\{^{\,\,\rho}_{\mu\,\nu}\}_g+V^\rho_{\phantom{\rho}\mu\nu},
\label{sol1}
\end{equation}
where $\{^{\,\,\rho}_{\mu\,\nu}\}_g$ is the Christoffel connection of the metric tensor $g_{\mu\nu}$ and $V^\rho_{\phantom{\rho}\mu\nu}$ is a tensor.
Consequently, the Ricci tensor of the affine connection $\Gamma^{\,\,\rho}_{\mu\,\nu}$ is given by~\cite{Scho}
\begin{equation}
R_{\mu\nu}(\Gamma)=R_{\mu\nu}(g)-\frac{2}{3}(S_{\nu:\mu}-S_{\mu:\nu})+V^\rho_{\phantom{\rho}\mu\nu:\rho}-V^\rho_{\phantom{\rho}\mu\rho:\nu}+V^\sigma_{\phantom{\sigma}\mu\nu}V^\rho_{\phantom{\rho}\sigma\rho}-V^\sigma_{\phantom{\sigma}\mu\rho}V^\rho_{\phantom{\rho}\sigma\nu},
\label{sol2}
\end{equation}
where $R_{\mu\nu}(g)$ is the Riemannian Ricci tensor of the metric tensor $g_{\mu\nu}$ and the colon denotes the covariant differentiation with respect to $\{^{\,\,\rho}_{\mu\,\nu}\}_g$.
Eq.~(\ref{Ein}) and symmetrized Eq.~(\ref{sol2}) give
\begin{equation}
R_{\mu\nu}(g)=\kappa\Theta_{\mu\nu}-\frac{\kappa}{2}\Theta_{\rho\sigma}g^{\rho\sigma}g_{\mu\nu}-V^\rho_{\phantom{\rho}(\mu\nu):\rho}+V^\rho_{\phantom{\rho}(\mu|\rho|:\nu)}-V^\sigma_{\phantom{\sigma}(\mu\nu)}V^\rho_{\phantom{\rho}\sigma\rho}+V^\sigma_{\phantom{\sigma}(\mu|\rho}V^\rho_{\phantom{\rho}\sigma|\nu)}.
\label{EMT}
\end{equation}
We also have~\cite{Scho}
\begin{equation}
Q_{\mu\nu}=-\frac{8}{3}(S_{\nu,\mu}-S_{\mu,\nu})+V^\rho_{\phantom{\rho}\rho\nu,\mu}-V^\rho_{\phantom{\rho}\rho\mu,\nu}.
\label{q1}
\end{equation}

Substituting Eq.~(\ref{sol1}) to Eq.~(\ref{field3}) gives
\begin{equation}
V^\mu_{\phantom{\mu}(\sigma\rho)}{\sf g}^{\sigma\nu}+V^\nu_{\phantom{\nu}(\rho\sigma)}{\sf g}^{\mu\sigma}-V^\sigma_{\phantom{\sigma}(\sigma\rho)}{\sf g}^{\mu\nu}=\Sigma_{\phantom{\mu}\rho\phantom{\nu}}^{\mu\phantom{\rho}\nu}.
\label{sol3}
\end{equation}
Its solution is
\begin{equation}
V^\rho_{\phantom{\rho}(\mu\nu)}=\frac{1}{2}(\Delta_{\phantom{\rho}\nu\phantom{\sigma}}^{\rho\phantom{\nu}\sigma}{\sf g}_{\mu\sigma}+\Delta_{\phantom{\rho}\mu\phantom{\sigma}}^{\rho\phantom{\mu}\sigma}{\sf g}_{\nu\sigma}-\Delta_{\phantom{\alpha}\gamma\phantom{\beta}}^{\alpha\phantom{\gamma}\beta}{\sf g}_{\mu\alpha}{\sf g}_{\nu\beta}{\sf g}^{\rho\gamma}),
\label{sol4}
\end{equation}
where
\begin{equation}
\Delta_{\phantom{\mu}\rho\phantom{\nu}}^{\mu\phantom{\rho}\nu}=\Sigma_{\phantom{\mu}\rho\phantom{\nu}}^{\mu\phantom{\rho}\nu}-\frac{1}{2}\Sigma_{\phantom{\alpha}\rho\phantom{\beta}}^{\alpha\phantom{\rho}\beta}g_{\alpha\beta}g^{\mu\nu}
\end{equation}
 and ${\sf g}_{\mu\nu}$ is the density reciprocal to ${\sf g}^{\mu\nu}$: ${\sf g}_{\mu\nu}{\sf g}^{\rho\nu}=\delta_\mu^\rho$.
Substituting Eq.~(\ref{Max2}) to Eq.~(\ref{field2}) and antisymmetrizing the indices $\mu$ and $\nu$ yield
\begin{equation}
V^\mu_{\phantom{\mu}[\sigma\rho]}{\sf g}^{\nu\sigma}-V^\nu_{\phantom{\nu}[\sigma\rho]}{\sf g}^{\mu\sigma}=\Omega_\rho^{\phantom{\rho}\mu\nu},
\label{field4}
\end{equation}
where
\begin{equation}
\Omega_\rho^{\phantom{\rho}\mu\nu}=\Pi_{\phantom{[\mu}\rho\phantom{\nu]}}^{[\mu\phantom{\rho}\nu]}-\frac{1}{3}\Pi_{\phantom{[\sigma}\sigma\phantom{\nu]}}^{[\sigma\phantom{\sigma}\nu]}\delta^\mu_\rho+\frac{1}{3}\Pi_{\phantom{[\sigma}\sigma\phantom{\mu]}}^{[\sigma\phantom{\sigma}\mu]}\delta^\nu_\rho
\end{equation}
is a traceless tensor density.
Consequently, we find
\begin{equation}
V^\rho_{\phantom{\rho}[\mu\nu]}=\frac{1}{2}(\Omega_\nu^{\phantom{\nu}\rho\sigma}{\sf g}_{\mu\sigma}-\Omega_\mu^{\phantom{\mu}\rho\sigma}{\sf g}_{\nu\sigma}-\Omega_\gamma^{\phantom{\gamma}\alpha\beta}{\sf g}_{\mu\alpha}{\sf g}_{\nu\beta}{\sf g}^{\rho\gamma}).
\label{sol5}
\end{equation}
Eqs.~(\ref{sol4}) and~(\ref{sol5}) give the tensor $V^\rho_{\phantom{\rho}\mu\nu}$.
If there are no sources, $\Pi_{\phantom{\mu}\rho\phantom{\nu}}^{\mu\phantom{\rho}\nu}=0$, the connection $\Gamma^{\,\,\rho}_{\mu\,\nu}$ depends only on the metric tensor $g_{\mu\nu}$ representing a free gravitational field and the torsion vector $S_\mu$ corresponding to the vectorial degree of freedom associated with the electromagnetic potential $A_\mu$~\cite{FK2}.

The purely metric formulation of gravitation is dynamically equivalent to the purely affine and metric--affine formulation, which can be shown by applying to $\textgoth{H}(\Gamma,{\sf g},Q)$ the Legendre transformation with respect to $\Gamma^{\,\,\rho}_{\mu\,\nu}$~\cite{FK3}.
This transformation defines the Lagrangian density in the momentum space $\textgoth{K}$:
\begin{equation}
\textgoth{K}=\textgoth{H}-\frac{\partial\textgoth{H}}{\partial\Gamma^{\,\,\rho}_{\mu\,\nu}}\Gamma^{\,\,\rho}_{\mu\,\nu}=\textgoth{H}+\frac{1}{2\kappa}\Pi_{\phantom{\mu}\rho\phantom{\nu}}^{\mu\phantom{\rho}\nu}\Gamma^{\,\,\rho}_{\mu\,\nu},
\label{Leg2}
\end{equation}
which satisfies: $d\textgoth{K}=\frac{\partial\textgoth{K}}{\partial{\sf g}^{\mu\nu}}d{\sf g}^{\mu\nu}+\frac{1}{2\kappa}\Gamma^{\,\,\rho}_{\mu\,\nu}d\Pi_{\phantom{\mu}\rho\phantom{\nu}}^{\mu\phantom{\rho}\nu}+\frac{\partial\textgoth{K}}{\partial Q_{\mu\nu}}dQ_{\mu\nu}$.
Accordingly, the momentum Lagrangian density $\textgoth{K}$ is a function of ${\sf g}^{\mu\nu}$, $\Pi_{\phantom{\mu}\rho\phantom{\nu}}^{\mu\phantom{\rho}\nu}$ and $Q_{\mu\nu}$, and the action variation is: $\delta S=\frac{1}{c}\delta\int\Bigl(\textgoth{K}({\sf g},\Pi,Q)-\frac{1}{2\kappa}{\sf g}^{\mu\nu}P_{\mu\nu}(\Gamma)-\frac{1}{2\kappa}\Pi_{\phantom{\mu}\rho\phantom{\nu}}^{\mu\phantom{\rho}\nu}\Gamma^{\,\,\rho}_{\mu\,\nu}\Bigr)d^4x=\frac{1}{c}\int\Bigl(\frac{\partial\textgoth{K}}{\partial{\sf g}^{\mu\nu}}\delta{\sf g}^{\mu\nu}+\frac{\partial\textgoth{K}}{\partial\Pi_{\phantom{\mu}\rho\phantom{\nu}}^{\mu\phantom{\rho}\nu}}\delta\Pi_{\phantom{\mu}\rho\phantom{\nu}}^{\mu\phantom{\rho}\nu}-\frac{1}{2\kappa}{\sf h}^{\mu\nu}\delta Q_{\mu\nu}-\frac{1}{2\kappa}{\sf g}^{\mu\nu}\delta P_{\mu\nu}-\frac{1}{2\kappa}P_{\mu\nu}\delta{\sf g}^{\mu\nu}-\frac{1}{2\kappa}\Pi_{\phantom{\mu}\rho\phantom{\nu}}^{\mu\phantom{\rho}\nu}\delta\Gamma^{\,\,\rho}_{\mu\,\nu}-\frac{1}{2\kappa}\Gamma^{\,\,\rho}_{\mu\,\nu}\delta\Pi_{\phantom{\mu}\rho\phantom{\nu}}^{\mu\phantom{\rho}\nu}\Bigr)d^4x$.
The variation with respect to ${\sf g}^{\mu\nu}$ yields the Einstein equations:
\begin{equation}
P_{\mu\nu}=2\kappa\frac{\partial\textgoth{K}}{\partial{\sf g}^{\mu\nu}}.
\label{Ham2}
\end{equation}
The variations with respect to $P_{\mu\nu}$ and $Q_{\mu\nu}$ can be transformed to the variation with respect to $\Gamma^{\,\,\rho}_{\mu\,\nu}$ by means of the Palatini formula and the variation of a curl, respectively, giving the field equations~(\ref{field1}).
Finally, the variation with respect to $\Pi_{\phantom{\mu}\rho\phantom{\nu}}^{\mu\phantom{\rho}\nu}$ gives
\begin{equation}
\Gamma^{\,\,\rho}_{\mu\,\nu}=2\kappa\frac{\partial\textgoth{K}}{\partial\Pi_{\phantom{\mu}\rho\phantom{\nu}}^{\mu\phantom{\rho}\nu}},
\label{Ham3}
\end{equation}
in accordance with Eq.~(\ref{Leg2}).

The analogous transformation in classical mechanics goes from a Hamiltonian $H(q^i,p^i)$ to a momentum Lagrangian $K(p^i,\dot{p}^i)=-f^j q^j-H(q^i,p^i)$.
The equations of motion result from Hamilton's principle written as $\delta\int(p^j\dot{q}^j+f^j q^j+K(p^i,\dot{p}^i))dt=0$.
The quantity $K$ is a Lagrangian with respect to $p^i$ because $p^j\dot{q}^j+f^j q^j$ is a total time derivative and does not affect the action variation.

If we define
\begin{equation}
C_\rho^{\phantom{\rho}\mu\nu}=\Sigma_{\phantom{\mu}\rho\phantom{\nu}}^{\mu\phantom{\rho}\nu}-\,^\ast\Gamma^{\,\,\,\,\mu}_{(\sigma\,\rho)}{\sf g}^{\sigma\nu}-\,^\ast\Gamma^{\,\,\,\,\nu}_{(\rho\,\sigma)}{\sf g}^{\mu\sigma}+\,^\ast\Gamma^{\,\,\,\,\sigma}_{(\sigma\,\rho)}{\sf g}^{\mu\nu},
\end{equation}
where the connection $\Gamma^{\,\,\rho}_{\mu\,\nu}$ depends on the source density $\Pi_{\phantom{\mu}\rho\phantom{\nu}}^{\mu\phantom{\rho}\nu}$ via Eq.~(\ref{con1}) or Eq.~(\ref{Ham3}), then the field equations~(\ref{field2}) and~(\ref{Max2}) can be written as
\begin{equation}
{\sf g}^{\mu\nu}_{\phantom{\mu\nu},\rho}=C_\rho^{\phantom{\rho}\mu\nu}.
\end{equation}
Accordingly, $\Pi_{\phantom{\mu}\rho\phantom{\nu}}^{\mu\phantom{\rho}\nu}$ can be expressed in terms of ${\sf g}^{\mu\nu}_{\phantom{\mu\nu},\rho}$ and the torsion tensor $S^\rho_{\phantom{\rho}\mu\nu}$.
Therefore a theory of gravitation with the connection as a dynamical variable contains in a natural way the first derivatives of the metric tensor, $g_{\mu\nu,\rho}$.
Consequently, the Christoffel symbols $\{^{\,\,\rho}_{\mu\,\nu}\}_g$, which are homogeneous linear functions of $g_{\mu\nu,\rho}$, correspond to generalized forces $f^i=\dot{p}^i$~\cite{FK3}.
The density $\textgoth{K}({\sf g},\Pi,Q)$ can be expressed in terms of $g_{\mu\nu,\rho}$ and $S^\rho_{\phantom{\rho}\mu\nu}$ instead of $\Pi_{\phantom{\mu}\rho\phantom{\nu}}^{\mu\phantom{\rho}\nu}$.
The resulting quantity, $\textgoth{K}(g,\partial g,S,Q)$, is a Lagrangian density for matter $\mathcal{L}_m$ in the purely metric formulation of gravitation with torsion.
Similarly, the tensor $P_{\mu\nu}(\Gamma)$ in Eq.~(\ref{Ham2}) can be expressed as $P_{\mu\nu}(\partial g,S)$ which can be decomposed into $R_{\mu\nu}(g)$ and terms with the torsion tensor that we denote as $R_{\mu\nu}(S)$.
As a result, Eq.~(\ref{Ham2}) can be written as
\begin{equation}
R_{\mu\nu}(g)=2\kappa\frac{\partial\textgoth{K}}{\partial{\sf g}^{\mu\nu}}-R_{\mu\nu}(S),
\label{Ham4}
\end{equation}
which corresponds to Eq.~(\ref{EMT}).
Thus the equivalence of the metric--affine and purely metric formulation of gravity with a torsionless connection~\cite{FK3} is also valid for a general connection with torsion~\cite{nonsym}.

If we can separate the Lagrangian $\textgoth{L}$ into the part that depends on the Ricci tensors and does not explicitly on the affine connection, and the part that depends on the connection and does not on the Ricci tensors, the tensor $\Theta_{\mu\nu}$ will represent the matter part that corresponds to the curvature, i.e. the electromagnetic field.
The terms in Eq.~(\ref{EMT}) that contain $V^\rho_{\phantom{\rho}\mu\nu}$ form the tensor which we denote as $\kappa(U_{\mu\nu}-\frac{1}{2}U_{\rho\sigma}g^{\rho\sigma}g_{\mu\nu})$.
The symmetric tensor $U_{\mu\nu}$ corresponds to the matter part that is generated by the connection, and is quadratic in the density $\Pi_{\phantom{\mu}\rho\phantom{\nu}}^{\mu\phantom{\rho}\nu}$.
Similarly, the right-hand side of Eq.~(\ref{Ham4}) can be written as $\kappa(T_{\mu\nu}-\frac{1}{2}T_{\rho\sigma}g^{\rho\sigma}g_{\mu\nu})$, bringing Eq.~(\ref{Ham2}) to the standard general-relativistic form.
The symmetric energy--momentum tensor $T_{\mu\nu}$ represents total matter: $T_{\mu\nu}=\Theta_{\mu\nu}+U_{\mu\nu}$.
The Bianchi identity $(R^{\mu\nu}(g)-\frac{1}{2}R(g)g^{\mu\nu})_{:\nu}=0$ yields the covariant conservation of this tensor: $T^{\mu\nu}_{\phantom{\mu\nu}:\nu}=0$.

The purely affine Lagrangian density for the Maxwell electromagnetic field was found by Ferraris and Kijowski~\cite{FK1}, using the symplectic formulation of particle dynamics and field theory by Tulczyjew~\cite{KT}.
This Lagrangian density is given by
\begin{equation}
\textgoth{L}_{EM}=-\frac{1}{4}\sqrt{-\wp}F_{\alpha\beta}F_{\rho\sigma}P^{\alpha\rho}P^{\beta\sigma},
\label{MaxA1}
\end{equation}
where $F_{\mu\nu}=A_{\nu,\mu}-A_{\mu,\nu}$ is the electromagnetic field tensor and $\wp=\mbox{det}P_{\mu\nu}$.
It is dynamically equivalent to the metric Lagrangian density for the Maxwell electromagnetic field, $\textgoth{H}_{EM}=-\frac{1}{4}\sqrt{-g}F_{\alpha\beta}F_{\rho\sigma}g^{\alpha\rho}g^{\beta\sigma}$.
This equivalence follows from the identity: $\sqrt{-\wp}F_{\alpha\beta}P^{\alpha\rho}P^{\beta\sigma}=\sqrt{-g}F_{\alpha\beta}g^{\alpha\rho}g^{\beta\sigma}$, which results from Eqs. (\ref{met1}), (\ref{Leg1}) and (\ref{Ham1}) applied to $\textgoth{L}_{EM}$ and $\textgoth{H}_{EM}$.
This equivalence is violated if we add to the expression~(\ref{MaxA1}) a term that depends on the Ricci tensor, e.g., the Eddington affine Lagrangian density for the cosmological constant, $\textgoth{L}_{\Lambda}=\frac{1}{\kappa\Lambda}\sqrt{-\wp}$~\cite{Niko2}.

The formal similarity between $F_{\mu\nu}$ and the second Ricci tensor $Q_{\mu\nu}$ (both tensors are curls) suggests that the purely affine Lagrangian density for the unified electromagnetic and gravitational fields is given by
\begin{equation}
\textgoth{L}_{EM}=-\frac{e^2}{4}\sqrt{-\wp}Q_{\alpha\beta}Q_{\rho\sigma}P^{\alpha\rho}P^{\beta\sigma},
\label{MaxA2}
\end{equation}
where $e$ has the dimension of electric charge~\cite{FK2}.
Without loss of generality, $e$ can be taken equal to the charge of the electron.
The dynamical equivalence between this Lagrangian and the Maxwell electrodynamics follows from the equivalence between the latter and $\textgoth{L}_{EM}$, since replacing $F_{\mu\nu}$ by $eQ_{\mu\nu}$ does not affect the algebraic relation of ${\sf g}^{\mu\nu}$ to $P_{\mu\nu}$ arising from Eqs.~(\ref{met1}) and~(\ref{Ham1}).
Eq.~(\ref{amet1}) gives
\begin{equation}
{\sf h}^{\mu\nu}=\kappa e^2\sqrt{-\wp}Q_{\alpha\beta}P^{\mu\alpha}P^{\nu\beta}.
\label{amet2}
\end{equation}
The torsion vector is related to the electromagnetic potential via Eq.~(\ref{q1}) and the correspondence relation $F_{\mu\nu}=eQ_{\mu\nu}$:
\begin{equation}
S_\nu=\frac{3}{8}(-\frac{A_\nu}{e}+V^\rho_{\phantom{\rho}\rho\nu}).
\label{tor}
\end{equation}
The gauge transformation $A_\nu\rightarrow A_\nu+\partial_\nu \lambda$, where $\lambda$ is a scalar function of the coordinates, does not affect Eq.~(\ref{q1}).

In order to write the Maxwell equations with sources in the purely affine formulation of gravity, we need to know how the affine Lagrangian for matter (independent of the curvature) depends on the connection.
The simplest choice is
\begin{equation}
\textgoth{L}_{m}=\rho u^\mu_{\phantom{\mu};\mu}=\rho(u^\mu_{\phantom{\mu},\mu}+\Gamma^{\,\,\mu}_{\nu\,\mu}u^\nu),
\label{mat1}
\end{equation}
where $\rho$ is a scalar density and $u^\mu$ is the local four-velocity of matter.
The four-velocity can be uniquely defined by the fluid configuration and its partial derivatives, without any knowledge of the metric~\cite{KM}.
Consequently, we find:
\begin{equation}
\Pi_{\phantom{\mu}\sigma\phantom{\nu}}^{\mu\phantom{\sigma}\nu}=-2\kappa\frac{\partial\textgoth{L}_{m}}{\partial \Gamma^{\,\,\rho}_{\mu\,\nu}}=-2\kappa\rho u^\mu\delta^\nu_\sigma
\label{mat2}
\end{equation}
and ${\sf j}^\mu=-\frac{1}{4}\kappa\rho u^\mu$.
Eqs.~(\ref{Max2}) and~(\ref{amet2}) yield
\begin{equation}
e^2(\sqrt{-\wp}Q_{\alpha\beta}P^{\nu\alpha}P^{\mu\beta})_{,\nu}=-\frac{1}{4}\rho u^\mu.
\label{MaxA3}
\end{equation}
Eq.~(\ref{MaxA3}) reproduces the Maxwell equations with sources, $\frac{1}{\sqrt{-g}}(\sqrt{-g}F_{\alpha\beta}g^{\nu\alpha}g^{\mu\beta})_{,\nu}=j^\mu=\frac{\rho_e c}{\sqrt{g_{00}}u^0}u^\mu$, where $j^\mu$ is the electromagnetic current four-vector and $\rho_e$ is the spatial density of electric charge~\cite{LL2}, if we identify $-\frac{\rho u^0}{4e}$ with $\frac{\rho_e c\sqrt{-g}}{\sqrt{g_{00}}}$.

For the matter Lagrangian~(\ref{mat1}) we find:
\begin{equation}
\Omega_\sigma^{\phantom{\sigma}\mu\nu}=0,\,\,\,\,V^\sigma_{\phantom{\sigma}[\mu\nu]}=0,\,\,\,\,\Sigma_{\phantom{\mu}\sigma\phantom{\nu}}^{\mu\phantom{\sigma}\nu}=\kappa\rho u^{(\mu}\delta^{\nu)}_\sigma,\,\,\,\,\Delta_{\phantom{\mu}\sigma\phantom{\nu}}^{\mu\phantom{\sigma}\nu}=\kappa\rho\Bigl(u^{(\mu}\delta^{\nu)}_\sigma-\frac{1}{2}u_\sigma g^{\mu\nu}\Bigr).
\label{spe}
\end{equation}
Accordingly, the torsion tensor depends only on the torsion vector, as in the case without sources.
Combining Eqs.~(\ref{sol4}), (\ref{sol5}) and~(\ref{mat2}) gives
\begin{equation}
V^\sigma_{\phantom{\sigma}\mu\nu}=\frac{\kappa\rho}{\sqrt{-g}}\Bigl(\frac{3}{4}u^\sigma g_{\mu\nu}-\frac{1}{2}u_{(\mu}\delta^\sigma_{\nu)}\Bigr),
\label{V}
\end{equation}
which after substituting to Eq.~(\ref{EMT}) yields
\begin{equation}
R_{\mu\nu}(g)=\kappa\Theta_{\mu\nu}+\frac{3}{8}\Bigl(\frac{\kappa\rho}{\sqrt{-g}}\Bigr)^2 u_\mu u_\nu-\frac{3}{4}\Bigl(\frac{\kappa\rho}{\sqrt{-g}}u^\sigma\Bigr)_{:\sigma}g_{\mu\nu},
\label{R}
\end{equation}
where $\Theta_{\mu\nu}=\frac{1}{4}g_{\mu\nu}F_{\alpha\beta}F^{\alpha\beta}-F_{\mu\rho}F_\nu^{\phantom{\nu}\rho}$.
The last term in Eq.~(\ref{R}) vanishes since ${\sf j}^\mu$ is covariantly conserved.
Finally, if we define the effective energy density $\epsilon$ and pressure $p$ via:
\begin{equation}
\epsilon+p\equiv\frac{3\kappa}{8}\Bigl(\frac{\rho}{\sqrt{-g}}\Bigr)^2,\,\,\,\,p\equiv\frac{3\kappa}{16}\Bigl(\frac{\rho}{\sqrt{-g}}\Bigr)^2 u_\sigma u^\sigma,
\label{ep}
\end{equation}
we arrive at the Einstein equations:
\begin{equation}
G_{\mu\nu}=\kappa\Bigl(\frac{1}{4}g_{\mu\nu}F_{\alpha\beta}F^{\alpha\beta}-F_{\mu\rho}F_\nu^{\phantom{\nu}\rho}+(\epsilon+p)u_\mu u_\nu-pg_{\mu\nu}\Bigr),
\label{equation}
\end{equation}
where $G_{\mu\nu}=R_{\mu\nu}(g)-\frac{1}{2}R(g)g_{\mu\nu}$ is the Einstein tensor of general relativity.

The Bianchi identities $G^{\mu\nu}_{\phantom{\mu\nu}:\nu}=0$ and $F_{[\mu\nu:\rho]}=0$ give the continuum form of the Lorentz equation of motion: $((\epsilon+p)u^\mu u^\nu)_{:\nu}-p^{,\mu}=F^{\mu\nu}j_\nu$.
Integrating it over space with $p=0$ leads to the Lorentz equation of a particle with mass $m\propto\epsilon$ and charge $q\propto\rho_e$.
Eqs.~(\ref{ep}) indicate that $\epsilon\propto\rho_e^2$, thus $m\propto q^2$, i.e. the mass of a particle is proportional to its intrinsic electromagnetic energy.
Therefore, the simplest purely affine Lagrangian for matter~(\ref{mat1}) generates the Einstein and Lorentz equations if {\em mass has electromagnetic origin}.
This conclusion may be crucial in generalizing quantum gravity in the Eddington purely affine picture~\cite{QG} to the case with the presence of matter and the electromagnetic field.

On larger scales, electric charge (either positive or negative) averages to zero, while mass (always positive) does not.
As a result, the continuum Lorentz equation averages to the geodesic equation, $u^\mu_{\phantom{\mu}:\nu}u^\nu=0$.
The equivalence of a purely affine gravity with general relativity, which is a metric theory, implies that the former is consistent with experimental tests of the weak equivalence principle~\cite{Wi}.

The unified field theory presented in this paper may be regarded as a mathematical exercise that attempts to combine gravitation and electromagnetism on the classical level.
A general affine connection has enough degrees of freedom to make it possible to describe the gravitational and electromagnetic fields.
In general relativity, the electromagnetic field and its sources are considered to be on the side of the matter tensor in the field equations, i.e. they act as sources of the gravitational field.
In unified field theory, the electromagnetic field obtains the same geometric status as the gravitational field.
While the ultimate goal is to geometrize material sources of both fields as well, in this paper we introduced matter through an extra Lagrangian containing non-geometrized matter variables.
We showed that a simple matter Lagrangian that contains two variables: a scalar density related to the electric charge density and the four-velocity of the continuous matter distribution, leads to the Lorentz force appearing as a consequence of the field equations for the particles whose masses are of purely {\em electromagnetic} origin.
However, there are massive particles in nature, like protons and neutrons, whose masses cannot be attributed only to the electromagnetic interaction.
Thus we must either modify the matter Lagrangian or extend the dependence of a purely affine Lagrangian on the curvature to the full Riemann tensor.

Finally, we emphasize the remarkable role of the electromagnetic field in the purely affine gravity.
Purely affine Lagrangians that depend explicitly on a general, {\em unconstrained} affine connection and the symmetric part of the Ricci tensor are subject to an unphysical constraint on the source density.
The inclusion of the second Ricci tensor, which is related to the electromagnetic field, in a purely affine Lagrangian {\em replaces} this constraint with the Maxwell equations and {\em preserves} the projective invariance of the Lagrangian without constraining the connection.

\end{document}